\documentclass[cameraready]{Interspeech}
\UseRawInputEncoding

\title{A Large-Scale Per-Speaker Analysis of Re-identification Risk in Speech Anonymization}

\author[affiliation={1}]{Orane}{Dufour}
\author[affiliation={1}]{Paul}{Magron}
\author[affiliation={2}]{Mickael}{Rouvier}
\author[affiliation={1}]{Emmanuel}{Vincent}

\address{
    $^1$ Universit\'{e}
 de Lorraine, CNRS, Inria, LORIA, F-54000 Nancy, France \\
    $^2$ LIA, Avignon University, F-84911 Avignon, France
}

\email{\{orane.dufour, paul.magron, emmanuel.vincent\}@inria.fr, mickael.rouvier@univ-avignon.fr}

\usepackage{comment}

\begin{document}

\maketitle

\begin{abstract}
Speech anonymization is commonly evaluated using average-case metrics such as the equal error rate, which can hide large disparities in re-identification risks across individuals. In this paper, we conduct a large-scale per-speaker privacy analysis using a linkability-based metric under a worst-case scenario. Nearly 5,000 speakers are evaluated across multiple anonymization systems, attacker architectures, and conversation lengths. While linkability scores are highly polarized at the speaker level, the sets of easy to re-identify and hard to re-identify speakers vary substantially across configurations. We show that no single factor explains speaker vulnerability. Instead, the re-identification risk emerges from the interaction between the attacker, the anonymizer, and the amount of available speech. These results challenge the notion of intrinsic speaker-level privacy risks and emphasize the need for evaluation protocols that are explicitly conditioned on the attacker and anonymizer.
\end{abstract}

\section{Introduction}
\label{sec:intro}
Speech conveys a multitude of personal information about the speaker such as biometric identity, age, gender, health condition, or emotional state~\cite{Kröger2020}. As such, the storage and processing of large speech datasets puts privacy protection at risk~\cite{Kröger2020, NAUTSCH2019441}. Existing data protection regulations, such as those outlined in the General Data Protection Regulation (GDPR)~\cite{gdpr2016} by the European Parliament, have led to the development of speech anonymization methods. These methods aim to remove speaker-specific attributes from a speech signal, while preserving a signal that remains usable for other speech processing tasks. The Voice Privacy Challenge (VPC), first introduced in 2020~\cite{Tomashenko_2022}, has played a central role in fostering progress by providing common baselines and benchmarks for anonymization systems. In this context, privacy is typically evaluated through an automatic speaker verification (ASV)~\cite{bai2021speakerrecognitionbaseddeep} system, also referred to as the attacker, whose task is to re-identify speakers by comparing anonymized trial utterances against enrollment utterances with known speaker identities. However, until the 2025 Voice Privacy Attacker Challenge~\cite{Tomashenko_2025}, limited attention was paid to the adversarial perspective. This latest edition highlighted how severely privacy risks had been underestimated once state-of-the-art ASV attackers were considered.

Still, the performance of ASV systems is most often reported using the equal error rate (EER)~\cite{tomashenko2024voiceprivacy2024challengeevaluation}, which only provides an average-case view of privacy risks. Such aggregate measures can obscure substantial variability across speakers, masking cases of specific individuals that are highly vulnerable to re-identification. To remain fair, privacy protection should not hinge on population averages, but instead ensure guarantees for all users. Therefore a more appropriate evaluation must adopt a worst-case perspective that explicitly accounts for individual exposure. In this direction, several works have proposed metrics that explicitly target worst-case leakage, such as the ZEBRA framework~\cite{ZEBRA} or GDPR inspired notions of linkability and singling-out~\cite{vauquier25_interspeech}. These approaches highlight the limitations of the EER, showing that it fails to capture the nuanced variations in the residual privacy risk under different attack conditions. While such metrics are indeed more suitable for analyzing speaker-level vulnerabilities than the EER, in practice their outcomes are typically reported as averages over the dataset, which obscures these vulnerabilities and prevents these metrics from being fully exploited for worst-case risk evaluation.

So far, the literature on the analysis of ASV scores remains relatively limited. In~\cite{williams2024anonymizing}, Williams and al. investigated speaker-level score distributions by identifying subpopulations with distinct behaviors, but only under ignorant and lazy-informed attack models and within the scope of the VPC, which includes only 40 test speakers. In~\cite{10447137}, the authors examined bias in a VPC baseline system by conducting subgroup analyses based on gender and dialect, revealing disparities across groups. The authors in~\cite{comprehensiveFramework} carried out similar experiments across multiple datasets and observed that disparities (e.g., gender) were not always present in all of them. They also report that  men were more protected in some cases and women in some others. Finally, \cite{youarewhatyousay} conducted a speaker-level study under a semi-informed as defined in~\cite{srivastava:tel-03674540} (worst-case) attacker against all three baseline anonymization systems, again restricted to the VPC test set. They showed that the anonymity is repeatedly compromised for certain speakers and, for some of these, the linguistic content alone suffices for re-identification.

While theses studies have shown that anonymization affects speakers unevenly, they mostly rely on subgroup analyses based on predefined labels which limits scalability and may overlook unknown factors influencing privacy risks. They also typically involve small datasets and limited diversity in ASV architectures. To address these limitations, in this paper we move beyond label-based subgrouping toward large-scale per-speaker analysis driven directly by attacker behavior. Building on~\cite{vauquier25_interspeech}, we conduct a systematic per-speaker evaluation under multiple anonymization strategies and attacker models, with the goal of disentangling the respective impacts of speaker identity, anonymization, and ASV architecture on the re-identification risk. Comparing speaker score distributions from diverse attack settings reveals that few speakers are consistently easy or hard to re-identify, and that attack and anonymization conditions strongly influence the individual privacy risk.

The rest of this paper is structured as follows. Section~\ref{sec:Methodology} presents our per-speaker evaluation framework. Section~\ref{sec:Exp} describes the data, attackers, and anonymizers. Section~\ref{sec:analysis} reports and discusses the results. Finally, Section~\ref{sec:analysis} concludes the paper.

\section{Methodology}
\label{sec:Methodology}


\subsection{Attacker strategy}
Within the speech anonymization framework, the attacker is modeled as an ASV system. This deep neural network, trained on a speaker identification task, learns to extract speaker-discriminative acoustic features from speech. The last hidden layer is used as speaker representation (speaker embedding) of the utterance. During an attack, embeddings are extracted from a set of trial utterances whose speaker identities the attacker aims to recover. Embeddings are also obtained from enrollment utterances for which the speaker identities are known. The attacker then compares pairs constituted of a trial embedding $x_\mathrm{test}$ and an enrollment embedding $x_\mathrm{enroll}$, with a similarity metric \(s(\cdot,\cdot)\), in order to identify potential matches. Several evaluation metrics can be used to determine whether a match occurs and to quantify the attacker's success. In this work, we adopt linkability\footnote{The singling-out metric~\cite{vauquier25_interspeech} was excluded from the experiments because no implementation was available at the time of writing this paper.} as our evaluation metric.

\subsection{Linkability}
As defined in~\cite{vauquier25_interspeech}, linkability refers to the probability that an attacker can associate a test speaker embedding with its corresponding enrollment embedding for the same speaker $i$, among a set of $N$ enrollment speakers. 
A single attempt of linkage is said to succeed if
\begin{equation}\label{eq:linkage_condition}
s\bigl(x_{\mathrm{test}}^{(i)},\,x_{\mathrm{enroll}}^{(i)}\bigr)
>
\max_{j\neq i} s\bigl(x_{\mathrm{test}}^{(i)},\,x_{\mathrm{enroll}}^{(j)}\bigr).
\end{equation}
The linkability metric is then defined as the probability of successful linkage across all test samples and all speakers:
\begin{equation}
{\Pr}_{x_{\mathrm{test}}^{(i)}}\left\{
s\bigl(x_{\mathrm{test}}^{(i)},x_{\mathrm{enroll}}^{(i)}\bigr)
>
\max_{j \ne i} s\bigl(x_{\mathrm{test}}^{(i)},x_{\mathrm{enroll}}^{(j)}\bigr)
\right\}.
\end{equation}
To compute linkability on our test set, we adopt the protocol introduced in \cite{vauquier25_interspeech}. For each test speaker in the trial set, the distance between the test embedding and a pool of \(N\) enrollment speakers is computed using cosine similarity. The number of enrollment speakers \(N\) takes 11 values, ranging from \(22{,}024\) (the maximum number of available speakers in our enrollment set) down to \(21\). These values are obtained by repeatedly dividing the pool size: starting from \(22{,}024\), we divide by two to obtain the next value, and continue this process until reaching \(21\). This ensures coverage of a wide range of values while preserving a reasonable computational cost. To reduce the influence of any particular selection of enrollment speakers, we perform five random draws of the enrollment pool for each test speaker. For every draw, enrollment embeddings are computed using all available utterances of the selected speakers.

The result depends on two parameters: the number of enrollment speakers \(N\) and the conversation length \(L\). The conversation length corresponds to the number of utterances used to compute the test speaker embedding. We consider three values: \(L = 1\), \(3\), and \(5\). This design enables us to quantify to what extent the amount of speech data available per speaker impacts the re-identification risk and also, as stated in \cite{Franzreb_evaluation}, to reduce the variance of the scores when \(L > 1\).

\subsection{Easy- and hard-to-link speakers}
Since our experiments focus on individual speakers, we do not report linkability as a function of $N$ over the entire dataset. Instead, for each trial speaker, we compute an average linkability score obtained over multiple linkage trials. This score is averaged over 55 tests per speaker, corresponding to 5 random draws of $N$ enrollment speakers for each of the 11 values of $N$. We repeat this process for each attack scenario involving a different pair of attacker/anonymizer and a different value of conversation length $L$, and we generate a per-speaker score distribution for each configuration.

We are interested in the speakers who obtained a high linkability in those distributions and want to know if they remain the same in each of them. To do so, for each distribution, we compute the 3rd quartile (Q3) value (i.e., the value below which 75\% of the scores fall) and we generate lists with speakers who obtained a linkability above this value for all utterances. For comparison purposes, we also compute the 1st quartile value (Q1), and generate lists of speakers with a linkability equal or inferior to this value. These are denoted \textit{easy-to-link} speakers and \textit{hard-to-link} speakers.
Then, we compute the intersection of all easy-to-link speakers lists across distributions (i.e., considering all combinations of attackers, anonymizers, and conversation length $L$). We also compute the union of such lists, and similarly for hard-to-link speakers.
Furthermore, to have a more granular analysis and to determine the influence of each variable (the attacker, the anonymizer, and the conversation length), we also compare each pair of speaker lists using the Jaccard similarity.
The Jaccard similarity between two speaker lists  $S_1$ and $S_2$ is defined as:
\begin{equation}
\text{Jaccard}(S_1, S_2) = \frac{|S_1 \cap S_2|}{|S_1 \cup S_2|},
\end{equation}
where $|S_1 \cap S_2|$ is the number of speakers common to both lists, and
$|S_1 \cup S_2|$ is the total number of unique speakers across the two lists. This measure reflects the proportion of shared items relative to the total size of the two combined sets.

\section{Experimental setup}
\label{sec:Exp}

This section presents our experimental protocol. Our code is available for a reproductibility purpose\footnote{https://github.com/OraneD/Speaker-Linkability}.

\subsection{Datasets}
Table~\ref{tab:datasets} describes the datasets used in our experiments. We use LibriSpeech~\cite{LibriSpeech} to train the ASV models (more specifically we use the train-clean-360 split), and CommonVoice's 11th English release (CV 11.0)~\cite{ardila2020commonvoicemassivelymultilingualspeech} as the test set. CV 11.0 is split in an enrollment set A and a trial set B. Every speaker of set B is present in set A, with disjoint utterances. We use the exact same split as in~\cite{vauquier25_interspeech}.

\begin{table}[t]
  \setlength{\tabcolsep}{0.55\tabcolsep}
  \centering
  \caption{Datasets used for training and evaluation.}
  \begin{tabular}{l c c c c}
    \toprule
    \textbf{Dataset} & \textbf{Speakers} & \textbf{Dur. (h)} & \textbf{Utterances} & \textbf{Split} \\
    \midrule
    LibriSpeech & 921 & 360 & 104,014 & Train \\
    CV 11.0 A & 22,024 & 323 & 234,945 & Test (enrol.) \\
    CV 11.0 B & 4,949 & 1,409 & 996,971 & Test (trials) \\
    \bottomrule
  \end{tabular}
  \label{tab:datasets}
\end{table}

\subsection{Anonymization systems}
As anonymization systems, we employ two\footnote{Baseline B4~\cite{panariello2024speakeranonymizationusingneural} was excluded due to its encoder being trained on one of the CommonVoice releases.} baseline systems of VPC 2025  both based on neural voice conversion: B3~\cite{B3_2023} and B5~\cite{B5}. B3 extracts the linguistic content using explicit phonetic transcriptions and performs speech resynthesis with pseudo-speaker embeddings generated by a generative adversarial network~\cite{meyer22b_interspeech, meyer2022anonymizingspeechgenerativeadversarial}. In contrast, B5 encodes linguistic content using vector-quantized bottleneck features extracted from a wav2vec~2.0-based \cite{baevski2020wav2vec20frameworkselfsupervised} speech recognition model, and synthesizes speech with a HiFi-GAN vocoder trained on a set of (real) target speakers. Among the  baselines, B5 achieves the highest performance.

Anonymization for both the training and test set is consistently applied at the utterance level which means that each utterance of a given speaker is anonymized with the a different target speaker. This ensures that the evaluation reflects speaker identity rather than memorization of a fixed target. As shown by~\cite{franzreb2025improvingspeakeranonymizationevaluations}, with speaker level anonymization, an attacker may learn to recognize that target. By assigning a different target speaker for each utterance we mitigate biases introduced by potentially vulnerable source-target pairs.

\subsection{Attackers}

In order to assess robustness under a worst-case threat model we deliberately employ a diverse set of attacker architectures. Indeed, differences in model capacity, inductive biases, and input features can affect which speakers are re-identified, so architectural diversity will allow us to determine whether vulnerabilities are model-specific. We employ three different attackers:
\begin{enumerate} 
    \item \textbf{ECAPA\footnote{https://github.com/Voice-Privacy-Challenge/Voice-Privacy-Challenge-2024/tree/main}} is the baseline attacker of the VPC 2025. It is a TDNN/x-vector style architecture~\cite{ECAPA} that integrates channel and context-adaptive modules to produce highly discriminative speaker embeddings.
    \item \textbf{WavLM ECAPA\footnote{https://github.com/deep-privacy/sidekit}} is the same as ECAPA architecture-wise, but it processes WavLM~\cite{chen_2022} input features instead of log Mel filterbank features.
    \item \textbf{ResNet\footnote{https://github.com/kiwano-toolkit/kiwano/}}uses residual connections to enable very deep convolutional networks that learn hierarchical spectral-temporal representations~\cite{He2016}. We consider the deeper ResNet-101 trained with the Kiwano toolkit~\cite{kiwano2026}, which offers increased capacity to model subtle speaker cues. 
\end{enumerate} 
Note that we ran additional experiments with two more attackers~\cite{arefeen2026dastdualstreamvoiceanonymization, Yakovlev_2024}. The results are consistent with those obtained using the 3 attackers detailed above, but we do not include these due to space constraints. Each attacker has been trained on LibriSpeech-train-clean-360 in a \textit{semi-informed} scenario, which means that they have been trained on data anonymized with the same anonymizer being attacked. Note however that the source-to-target speaker mapping remains unknown by the attacker since it is done randomly. As a preliminary experiment, we evaluated each attacker following the 2025 VPC's evaluation protocol and computed the EER on LibriSpeech-test for both B3 and B5 anonymizers, in the semi-informed setting. Results match state-of-the-art performances for both WavLM ECAPA and ResNet, ensuring that the attackers are sufficiently strong for the subsequent analysis. Note that while the VPC 2025 winner~\cite{winnerVPC} would have been a natural choice for this study, no implementation is available thus we do not consider it here.

\section{Results}
\label{sec:analysis}
 
\subsection{Per-speaker score distributions}
\label{subsec:distributions}

\begin{figure*}[t]
    \centering
    \includegraphics[width=0.99\textwidth]{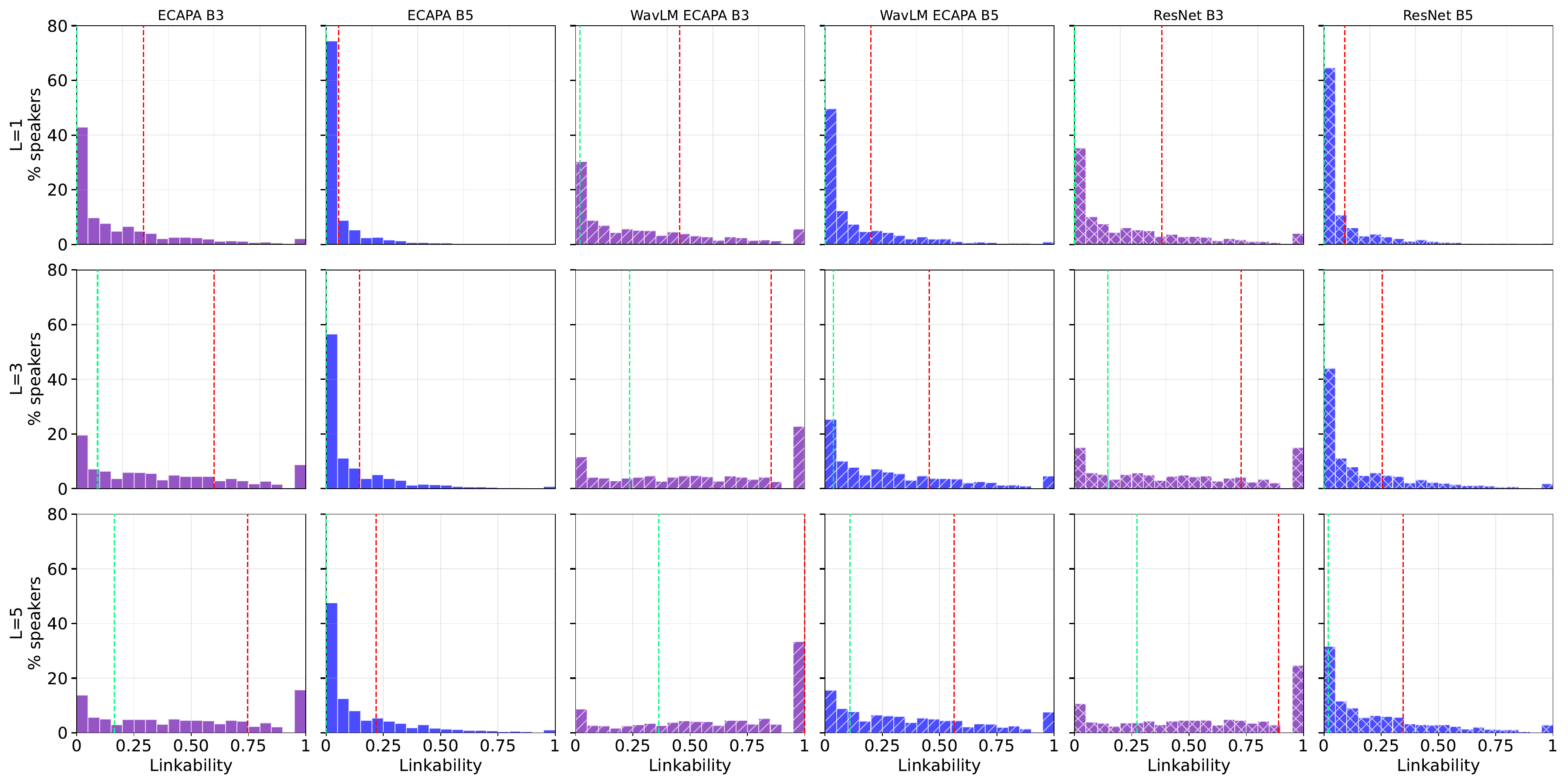}
    \caption{Distributions of the average linkability score obtained by all test speakers for all sets of attacker, anonymizer and conversation length $L$. Each line corresponds to a value of $L$ and each column to a pair of attacker and anonymizer. The purple and blue bars indicate an attack performed on B3 and B5, respectively. Dashed bars indicate the WavLM ECAPA attacker while crossed ones indicate ResNet. The green and red vertical dashed lines indicate Q1 and Q3, respectively.}
    \label{fig:distribution}
\end{figure*}

Figure~\ref{fig:distribution} shows the 18 speaker score distributions (2 anonymization systems $\times$ 3 attackers $\times$ 3  conversation length $L$). As shown in the figure, the WavLM ECAPA attacker consistently outperforms ResNet and ECAPA, and B3 is systematically easier to attack than B5. When $L=1$ most speakers obtain a linkability close to 0, whereas the opposite occurs when $L=5$ where most speakers reach a linkability of 1, meaning they were successfully linked to themselves in all 55 attempts. This confirms that increasing the conversation length drastically strengthens the attacker. Across all architectures and configurations, we observe a highly polarized distribution of speaker linkability: a substantial proportion of speakers achieve zero linkability, while another large group exhibits near-perfect linkability. This could reveal the presence of distinct speaker clusters, corresponding to inherently linkable versus inherently unlinkable speakers, highlighting a large heterogeneity in speaker vulnerability. However, when looking at the intersections and unions of easy- and hard-to-link speakers in the 18 different configurations, we observe that the identities of the speakers at both end of the distributions vary a lot. Indeed, considering the intersections as described in Table~\ref{tab:speaker_lists}, only 5 speakers are easy-to-link consistently across the 18 distributions, and only 166 of them are consistently hard-to-link. However, unions, for both easy- and hard-to-link speakers are very high: 86.9\% of the 4,949 test speakers are at least considered once as easy-to-link and 92.4\% of them are at least once hard-to-link.

\begin{table}[t]
\centering
\caption{Number of speakers in the intersection and union of easy- and hard-to-link speakers across the 18 distributions.}
\label{tab:speaker_lists}
\begin{tabular}{lcc}
\toprule
\textbf{Speakers} & \textbf{Intersection} & \textbf{Union} \\
\midrule
Easy-to-link  & 5  & 4,300 \\
Hard-to-link  & 166 & 4,574 \\
\bottomrule
\end{tabular}
\vspace{-8pt}
\end{table}

These results indicate that speaker vulnerability is strongly influenced by multiple system-related factors, namely the ASV architecture of the attacker, the anonymization model being attacked, and the value of the conversation length $L$. Importantly, these results suggest that vulnerability cannot be explained solely by intrinsic speaker characteristics, but also depends on the attack and anonymization configurations.

\subsection{Similarities between distributions}
\label{subsec:similarities}
\begin{figure}[t]
\vspace{-4pt}
    \centering
    \includegraphics[width=\columnwidth]{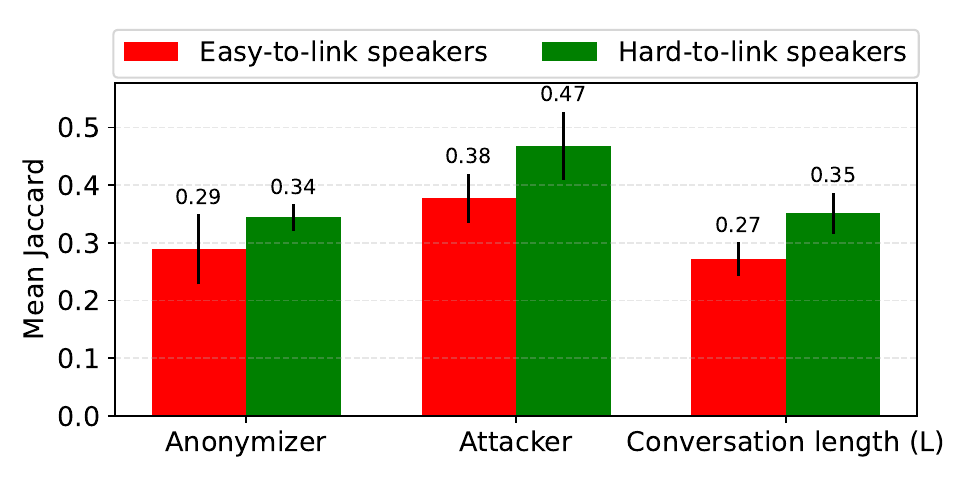}
    \caption{Mean Jaccard similarities for easy- and hard-to-link speakers. Black lines are standard deviations and the x-axis indicates the varying factor, while the other two are fixed.}
    \label{fig:Jaccard}
\vspace{-1em}
\end{figure}

To quantify the respective impact of the anonymizer, the ASV architecture, and the conversation length, on the linkability score distributions, we compute the Jaccard similarity between pairs of lists for both easy- and hard-to-link speakers. In each comparison, only one factor varies while the other two are kept fixed, isolating its effect on list composition. For each factor, we report the average Jaccard similarity across all such comparisons in Figure~\ref{fig:Jaccard}, for easy- and hard-to-link speakers separately.

First, we observe that the average Jaccard similarity never exceeds 0.47 for any group of speakers. This indicates that, regardless of the variable considered, the overlap between easy- or hard-to-link speaker lists is limited. In other words, changing any single variable leads to substantial changes in which speakers are considered easy or hard to link, suggesting that no single factor fully explains speaker vulnerability.
Second, Figure~\ref{fig:Jaccard} allows us to compare the relative impact of the different variables on the linkability score distributions. A higher Jaccard similarity indicates a smaller impact, as the identities of the speakers in the lists are more consistent across configurations.
\begin{itemize}
    \item \textbf{Attacker architecture:} this variable yields the highest Jaccard similarities (approximately 0.39 for easy-to-link speakers and 0.47 for hard-to-link speakers). This indicates that the composition of the easy- and hard-to-link speaker lists is more stable across different attacker architectures than across changes in the other variables, suggesting that the attacker architecture has the smallest impact on the distributions.
    \item \textbf{Anonymizer:} The anonymizer has a stronger influence than the attacker, with similarities around 0.29 and 0.34 for easy- and hard-to-link speakers respectively.

    \item \textbf{Conversation length:} The conversation length \(L\) yields Jaccard similarities very close to the anonymizer indicating that they have a comparable impact on the distributions.

\end{itemize}
Note that the differences in mean Jaccard similarities remain small, the largest gap being approximately 0.13. In particular, while the attacker architecture exhibits higher Jaccard similarities and can therefore be identified as having the smallest impact on the linkability score distributions, the anonymizer and the conversation length $L$ show very close mean values.

Moreover, the standard deviations associated with the anonymizer and the conversation length $L$ are of the same order of magnitude as the gap between their respective mean Jaccard similarities between easy- and hard-to-link speakers. This indicates that the observed differences between these two variables are not sufficiently pronounced to support a strict ordering. As a result, rather than establishing a definitive ranking, these results suggest that the anonymizer and the conversation length $L$ have a comparable impact on the linkability score distributions.

\section{Conclusion}
\label{sec:CONCLUSION}

We conducted the first large-scale per-speaker privacy risk analysis across multiple anonymization and attacker architectures and 
showed that the privacy risk does not depend solely on speakers. Instead, vulnerability to re-identification emerges from the interaction of the attacker's ASV architecture, the anonymization system, and the amount of speech 
for each test speaker. 
These results raise a fundamental challenge for privacy evaluation. 
The privacy risk 
under a given anonymization and attack configuration does not transfer across threat models, 
highlighting the need for context-dependent privacy guarantees.
Future work will focus on defining an \emph{a priori} re-identification risk~\cite{bousquet22_interspeech} conditioned on a fixed anonymization and attacker pair. We plan to explore 
reliability measures of ASV speaker embeddings as a means to estimate the privacy risk without explicit score-based evaluation. This direction is promising for more predictive and user-oriented privacy guarantees.

\clearpage
\newpage
\section{Acknowledgments}
This work was supported by the French Agence Nationale de la Recherche via the SpeechPrivacy project (ANR-23-CE23-0022). It was provided with computer and storage resources by GENCI at IDRIS thanks to the grant 2025-AD011015838R1 on the supercomputer Jean Zay's V100 partition.
\section{Use of Generative AI Disclosure}
We acknowledge the ISCA policy regarding the use of generative AI tools. The authors declare that generative AI tools have been used solely to correct grammar. No such tools were used to write significant parts of this manuscript.
\bibliographystyle{IEEEtran}
\bibliography{mybib}

\end{document}